\documentclass[12pt]{article}

\usepackage[utf8]{inputenc}
\usepackage{times}
\usepackage{xcolor}
\usepackage[pdftex]{graphicx}
\usepackage{float}
\usepackage{lineno}
\usepackage{epstopdf}
\usepackage[numbers]{natbib}
\input epsf
\newenvironment{sciabstract}{%
\begin{quote} \bf}
{\end{quote}}

\title{Mutual benefits of social learning and algorithmic mediation for cumulative culture}

\author{Agnieszka Czaplicka$^{1,2,*}$, Fabian Baumann$^{1,3}$, Iyad Rahwan$^{1}$\\
\\
\normalsize{$^{1}$Center for Humans and Machines, } \\ 
\normalsize{Max Planck Institute for Human Development, Lentzeallee 94, Berlin 14195, Germany.}\\
\normalsize{$^{2}$Faculty of Physics, Warsaw University of Technology, Warszawa, Poland.}\\
\normalsize{$^{3}$Department of Biology, University of Pennsylvania, Philadelphia, PA 19104.}\\
\normalsize{$^{*}$Corresponding author: agaczapl@gmail.com}
}

\date{}

\begin{document}

\baselineskip24pt

\maketitle 
\begin{sciabstract}
The remarkable ecological success of humans is often attributed to our ability to develop complex cultural artifacts that enable us to cope with environmental challenges. The evolution of complex culture (cumulative cultural evolution) is usually modeled as a collective process in which individuals invent new artifacts (innovation) and copy information from others (social learning). This classic picture overlooks the growing role of intelligent algorithms in the digital age (e.g., search engines, recommender systems, large language models) in mediating information between humans, with potential consequences for cumulative cultural evolution.
Building on a previous model, we investigate the combined effects of network-based social learning and a simplistic version of algorithmic mediation on cultural accumulation. We find that algorithmic mediation significantly impacts cultural accumulation and that this impact grows as social networks become less densely connected. Cultural accumulation is most effective when social learning and algorithmic mediation are combined, and the optimal ratio depends on the network’s density.
This work is an initial step towards formalising the impact of intelligent algorithms on cumulative cultural evolution within an established framework. Models like ours provide insights into mechanisms of human-machine interaction in cultural contexts, guiding hypotheses for future experimental testing.

\end{sciabstract}  
\section{Introduction}

Humans' ability to create complex cultural artifacts such as languages, scientific theories, art, and technology (including physical artifacts) is often seen as one of the most important aspects of our success as a species \cite{boyd1996culture,tomasello2009cultural}.  
An important driving force behind this cultural accumulation is the ability of humans to learn socially, i.e., copy information from their peers \cite{boyd1996culture,henrich2016secret,laland2017darwin,kolodny2015evolution}.
Social learning is a highly adaptive process, influenced by cognitive biases that can manifest as different social learning strategies ranging from copying successful individuals (payoff bias) \cite{henrich2001evolution,thompson2022complex,jimenez2019prestige}, to frequent cultural artifacts (context bias) \cite{hahn2003drift,herzog2004random}, to specific information (content bias) \cite{mesoudi2006bias}.  
At the population level, social learning is a necessary condition for cumulative cultural evolution (or cultural accumulation), a process that describes the emergence of increasingly complex cultural artifacts through a ratchet effect: By preserving important information over time—i.e., from generation to generation—social learning allows collectives to produce cultural artifacts that lie outside of the scope of individuals \cite{mesoudi2018cumulative,Henrich2004}.  
Yet the cultural accumulation can also be limited by factors such as population size \cite{Henrich2004}, the structure of the underlying social network \cite{derex2016partial}, or the social learning strategies that humans use \cite{miu2024refinement}.

In the digital age, cultural accumulation is increasingly shaped by algorithmic mediation \cite{brinkmann2023machine}. For the purposes of this paper, we broadly define algorithmic mediation as a collection of processes through which intelligent systems aggregate and redistribute information on an unprecedented scale, thereby impacting cultural accumulation. These algorithms include recommender systems that influence consumer behavior \cite{roy2022systematic}, content-ranking mechanisms on social media platforms \cite{cotter2017explaining}, and search engines like Google \cite{page1999pagerank}. More recently, large language models (LLMs) (e.g., GPT, BERT, Gemini, and others) have emerged as transformative tools, reshaping how we acquire knowledge and develop skills. Unlike earlier recommender systems, which primarily tackled e-commerce challenges by focusing on consumer preferences, LLMs are likely influencing a far broader range of domains that are central to cultural accumulation. These include generating ideas \cite{si2024can}, facilitating problem-solving \cite{wei2022chain}, assisting with technical tasks such as coding \cite{kazemitabaar2023novices}, and offering personalized explanations in fields like mathematics and science \cite{chevalier2024language}.
By synthesizing vast repositories of information and tailoring information to individual needs, algorithmic systems have indeed the potential to impact cumulative cultural evolution by enhancing innovation rates across domains, including technology and basic science \cite{cockburn2018impact}, reinforcing filter bubbles \cite{pariser2011filter}, and limiting the diversity of information shared \cite{mcpherson2001birds}.
Recognizing these impacts, previous theoretical and empirical works on socio-technological systems have mainly focused on the impact of algorithms on the diversity of content \cite{10.1145/3589334.3645713}, the formation of opinions \cite{perra2019modelling,lanzetti2023impactrecommendationsystemsopinion}, and the spread of misinformation \cite{guess2023social}.  
However, the question of how algorithmic mediation may affect cultural accumulation received much less attention.

To close this gap, we build on---and extend---an established model of cultural evolution \cite{10.1371/journal.pone.0018239}, where algorithmic mediation is conceptualized as a mechanism that aggregates and redistributes cultural traits on a collective level. By integrating social learning and algorithmic mediation in a single modeling framework, we are able to capture the dynamics induced by the interaction of localized peer influence (social learning) and global information redistribution (algorithmic mediation).
Our results shed light on an emerging trade-off that humans face in the digital age: cultural accumulation is neither optimal through pure social learning nor through purely algorithmically mediated information. Instead, both types of learning are most powerful when combined, where the optimal ratio depends on the underlying social network.  
This insight underscores the need to examine algorithmic mediation and social learning as interconnected forces shaping cultural evolution in the digital age, rather than as isolated processes.

\section{Model}
We build on prior theoretical work that explored how cultural accumulation is influenced by various aspects of individual learning, including social learning strategies and the costs associated with learning \cite{10.1371/journal.pone.0018239}. First, we extend this model by incorporating an explicit social network structure.
In a second step, we extend the model by a simple algorithm that mediates information exchange among agents.

The individual-based model presented in \cite{10.1371/journal.pone.0018239} simulates a dynamic process that unfolds over generations, where each "time step" represents the simultaneous birth and death of an entire generation of $N$ agents. In this framework, agents first copy information from the previous generation and then, after acquiring all available knowledge, begin to innovate. In this work, we deviate from this approach and modify the model by re-scaling the dynamics: at each time step, a single cultural trait is acquired either through copying or innovation \cite{Rendell208}. This modification enables us to examine the interplay between different mechanisms of information transmission—social learning and algorithmic mediation—and their implications for population-scale cultural accumulation. 

Below, we describe our modeling approach and provide additional implementation details in the Supplemental Material (SM).

\subsection{Cultural space}
We consider a space of cultural traits (or \textit{cultural space}) that defines how new traits (innovations) arise based on existing ones. 
In line with previous works, the cultural space consists of several cultural branches (or paths), and traits are functionally dependent, reflecting the cumulative nature of human culture \cite{derex2016partial,smolla2019cultural,10.1371/journal.pone.0018239,derex2018divide}.
 
The cultural space is structured as follows.
Each trait $x_l$ is defined by two components: the cultural branch $x$ it belongs to and its complexity level $l$, see Fig.~\ref{fig:scheme}A. 
Cultural branches $x$ model the fact that cultural evolution often follows multiple parallel paths \cite{smolla2019cultural,derex2016partial}, and levels $l$ model the functional dependency and increasing complexity of cultural traits built on top of each other. 
We consider a finite set of possible branches, i.e., $x=1,...,X$ and assume that cultural complexity levels $l$ are unbounded, i.e. $l=1,2,...,\infty$, i.e. cultural artifacts can be modified infinite number of times.
The functional dependency of the traits  $x$ additionally requires that the traits are acquired in ascending order, starting with the lowest level $l=1$, or more generally: an agent that wants to acquire a trait on level $l$ must possess a trait on level ${l-1}$.
Each cultural trait $x_l$ is characterised by an inherent quality score, or payoff, $z(x_l)$, and we assume that the extent to which the traits are beneficial to agents increases with $z$ \cite{kolodny2015evolution,derex2016partial,boyd1988culture,lazer2007network}.
Note, however, that we do not link an individual's probability to survive to their payoff. 
Instead, cultural selection happens through selective social learning, where agents \textit{copy from} agents with high payoffs \cite{boyd1988culture,henrich2003evolution}.
Motivated by empirical research on scientific \cite{park2023papers} and technological \cite{youn2015invention} breakthroughs and previous studies on cultural evolution \cite{kolodny2015evolution}, we assume that high-quality (i.e. high-payoff) innovations are rare among many low-quality attempts to innovate. 
Hence, we draw the payoffs of newly innovated items, $z(x)$, from an exponential distribution, and subsequently square, double, and finally round values to an integer value \cite{Rendell208}. 
This procedure results in a few traits with a payoff of around $z=50$, and roughly half of the traits with $z=0$ \cite{10.1371/journal.pone.0018239}.

The distinction between complexity levels $(l)$ and payoffs $(z)$ in the model captures the relationship between the accumulation of culture (i.e. knowledge, skills, or innovations) and its practical utility in real-world cultural dynamics. For instance, consider the development of a new scientific method (skill) or technological tool (physical artifact). The complexity of a tool---reflected in its levels---might represent the number of iterative modifications and refinements over time, such as the progression from early calculators to modern computers. However, not all refinements enhance utility or payoff. Some modifications, like aesthetic improvements or added features, may increase the complexity of a tool without necessarily maximizing its functional effectiveness or societal impact. Similarly, in scientific endeavors, researchers may pursue increasingly complex methodologies that yield incremental theoretical insights, but only some of these contribute substantially to practical applications or yield high-impact results capable of shifting the paradigm within a discipline. This distinction between levels and payoffs underscores the importance to disentangle how cumulative cultural processes lead to either optimal or suboptimal outcomes, depending on the social and algorithmic mechanisms mediating cultural transmission.

\subsection{Dynamics}
We consider a population of $N$ agents that are embedded in a social network. 
Initially, agents are {naive}, i.e. they do not possess any cultural traits, and their individual effort budget is $B$. A finite effort budget $B$ models the fact that individuals have limited resources, which also constrains cumulative cultural evolution at the collective level.
Over time, agents acquire cultural traits either by copying existing traits (learning) or by discovering new ones (innovation), with each action incurring a cost.
The dynamics of a focal agent $i$ at time $t$, characterized by the highest complexity level $l_i$ of their traits and their remaining resources $B(t)$, can be modeled as a cycle consisting of the following three events:

\begin{description}
    \item[$(i)$ \textit{Death-birth process}.] With probability $q=1/N$ agent $i$ is replaced by a naive agent, $i'$, which corresponds to the replacement of on average one agent per time step. The initial effort budget of agent $i'$ is $B$ and they do not possess any cultural traits. 
    Agent $i'$ takes the network position of agent $i$. 
    \item[$(ii)$ \textit{Social learning}.] Agent $i$ selects their neighbour $j$ with the highest cumulative payoff \cite{thompson2022complex}. 
    If agent $j$ is at a higher complexity level $l$ than agent $i$, i.e. $l_j>l_i$, agent $i$ copies the $(l_i+1)$--stage trait of agent $j$, and the learning cost $C_{s}$ is subtracted from agent $i$'s budget. If copying is not successful, either because the resources of agent $i$ are not sufficient ($B(t)<C_s$) or because the level of the highest trait of agent $j$ is too low, i.e. $l_j<l_{i+1}$, the learning cost $C_s$ is not deducted from the agent's budget. Agent $i$ then continues with the innovation step (see below). If the social learning step was successful, the innovation step is skipped. Social learning is schematically depicted in Fig.~\ref{fig:scheme}C. We mainly focus on payoff-biased learning as described above, however, in the SM we also show results for other social learning strategies.
    \item[$(iii)$ \textit{Innovation}.] If social learning was unsuccessful, and agent $i$ has a sufficient budget  ($B(t)\geq C_i$), they will proceed with innovation. 
    First, agent $i$ randomly selects a cultural branch $x$ and attempts to innovate a trait on level $l_i+1$.  
    If the trait is viable, i.e. its payoff $z(x_{l_i+1})>0$, the innovation step is successful and agent $i$ acquires the trait. Note that unsuccessful innovation concludes the time step for agent $i$ without acquiring a new item.     
    Regardless of the success of the innovation step, the cost of innovation $C_i$ is subtracted from agent $i$'s effort budget. Innovation is schematically depicted in Fig.~\ref{fig:scheme}B.
\end{description}

To analyze cultural accumulation in the model, we use three distinct measures. First, we calculate the mean payoff of all agents in the population, defined as 
$\bar{Z}(t) = \frac{1}{N} \sum\limits_i Z_i(t)$,
where $Z_i$ represents the cumulative payoff of agent $i$, given by $Z_i = \sum\limits_{l=1}^{l_\mathrm{max}} z(x_l)$.
Here, $z(x_l)$ denotes the payoff associated with trait $x_l$, and the summation is over the complexity levels $l$, reflecting the model's dynamics: each agent can acquire at most one trait per complexity level. The mean payoff $\bar{Z}(t)$ serves as a proxy for the \textit{performance} of the population, indicating the overall utility derived from the accumulated traits. 
Second, we count the total number of distinct traits created during the process. This measure provides an indicator of the \textit{diversity} of traits in cultural accumulation. 
Finally, we track the highest levels of traits reached by agents, which assess the average \textit{cultural complexity} within the population. 
These three measures---performance, diversity, and complexity---together offer a comprehensive framework for analyzing the dynamics of cultural accumulation in the model.

Note that we assume that the cost of acquiring a single trait depends only on the action performed, i.e., learning or innovation, but is independent of other factors, including the payoff $z$ and the complexity level $l$ of the trait.
Furthermore, we follow the intuition that it is easier to copy something from a peer than to innovate it, i.e. we assume $C_s<C_i$ \cite{10.1371/journal.pone.0018239}.

\subsection{Algorithmic mediation}

To investigate how algorithmic mediation affects cultural accumulation and the interplay between social and algorithm-mediated learning, we extend the model described in the previous section.  
While still embedded in a social network, agents can now also learn cultural traits through algorithmic mediation. We implement this as a 'perfect' algorithm with full knowledge of all cultural traits innovated so far, including their development levels $(l)$ and payoffs ($z$), enabling personalization of information.  
Learning through algorithmic mediation is assumed to be less costly than social learning ($C_r < C_s$). We implement the balance between social and algorithmic learning as a stochastic process: At each time step, an agent copies a trait via algorithmic mediation with probability $r$ or learns socially with probability $(1-r)$.

\begin{figure}[!ht]
    \centering
    \includegraphics[width=\textwidth]{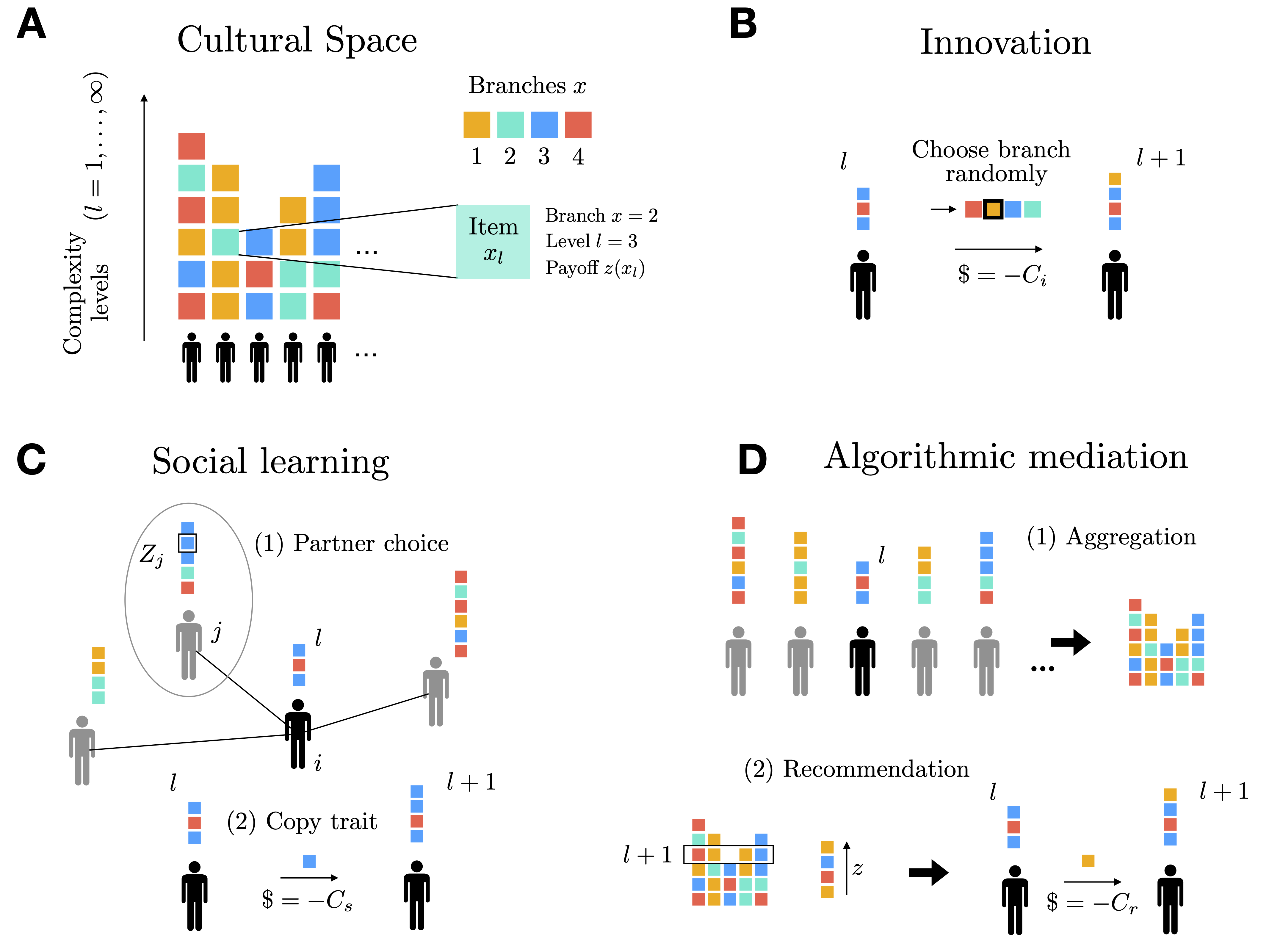}
\caption{Schematic depiction of key model components. A: Open-ended cultural space depicted with four branches ($X=4$), and an infinite number of (complexity) levels $l$. 
B: Agents innovate a new item on a randomly chosen branch on a level $l+1$. C: Social learning is modeled as a two-step process. (1) Agents choose their neighbour $j$ with the highest cumulative payoff $Z_j$. (2) If agent $j$ possesses an item on a sufficient level ($l+1$), the item is copied by agent $i$ and the cost $C_s$ is paid. D: Algorithmic mediation is also modeled as a two-step process. (1) The algorithm aggregates all existing cultural traits and sorts traits on level ($l+1$) according to their payoffs $z$. (2) Agent $i$ then receives a recommendation of the trait with the highest payoff.}
    \label{fig:scheme}
\end{figure}

\begin{description}
    \item[ \textit{Algorithmic recommendation} (with probability $r$).] In every time step, the algorithm aggregates all cultural traits and selects only those on level $l_i\!+\!1$ (personalization). After sorting those $l_i\!+\!1$--level traits in descending order (according to their payoffs) it recommends the highest-payoff trait, which is acquired by agent $i$ and the cost $C_r$ is deducted from their budget.
    Note that when no suitable traits can be recommended, or if agent $i$ does not have enough resources ($B(t)<C_r$), the recommendation fails, the agent does not pay a cost, and continues with the innovation step. When the recommendation is successful, agent $i$ does not attempt to innovate. Algorithmic mediation is schematically depicted in Fig.~\ref{fig:scheme}D.    
    \item[ \textit{Social learning} (with probability $1-r$).] Agent $i$ attempts to copy a cultural trait from their neighbours, as described in the previous section.
\end{description}

The order of events in each time step is identical to the one described in the previous section, i.e., consisting of $(i)$ the death-birth process, $(ii)$  learning (either socially or through algorithmic mediation), and $(iii)$ innovation. 
The different model components are schematically depicted in Fig.~\ref{fig:scheme}. Details of the implementation are presented in the SM.

While idealized, our model of algorithmic mediation broadly mirrors the capabilities of advanced algorithmic systems, such as large language models (LLMs), which tailor information delivery to align with user needs in order to maximize utility, although their effectiveness remains constrained by current methodological and technological limitations. For instance, reinforcement learning from human feedback, as employed in systems like ChatGPT, enables the prioritization of useful, accurate, and contextually relevant information. While this process is not directly analogous to biological fitness proxies in traditional cultural evolution models, it effectively captures the adaptive refinement and redistribution of cultural traits---including knowledge and skills---within an algorithmically mediated environment.  

\section{Results}
If not indicated otherwise, we present results for populations of $N=100$ agents that are embedded in a random social network, where each pair of individuals is connected with probability $p$. Hence, each agent has on average $k=p(N-1)$ neighbours. The effort budget of agents is set to $B=1000$, the costs for acquiring cultural traits are fixed at $C_i=10$, $C_s=5$, and $C_r=1$, and we assume $X=100$ cultural branches. The SM contains a sensitivity analysis of the model with respect to these parameters.

Due to the finite effort budget $B$ and the low probability of the death-birth process $q$, the system reaches a quasi-stationary state regardless of the remaining parameters. 
In the following, we will thus focus on "maximally achievable values",  i.e., population mean values at the stationary state for performance ($\bar{Z}$), diversity ($\bar{T}$), and complexity ($\bar{L}$). For brevity, we will refer to these values as "mean group values" and mark them with a subscript $_\mathrm{max}$. 
Each simulation continues for $t=20N$ time steps and the last $200$ time steps are used to compute the mean group values. 
Our results are reported as averages over $200$ realizations.

\subsection{Network-based social learning}
\begin{figure}[!ht]
    \centering
   \includegraphics[width=0.45\columnwidth]{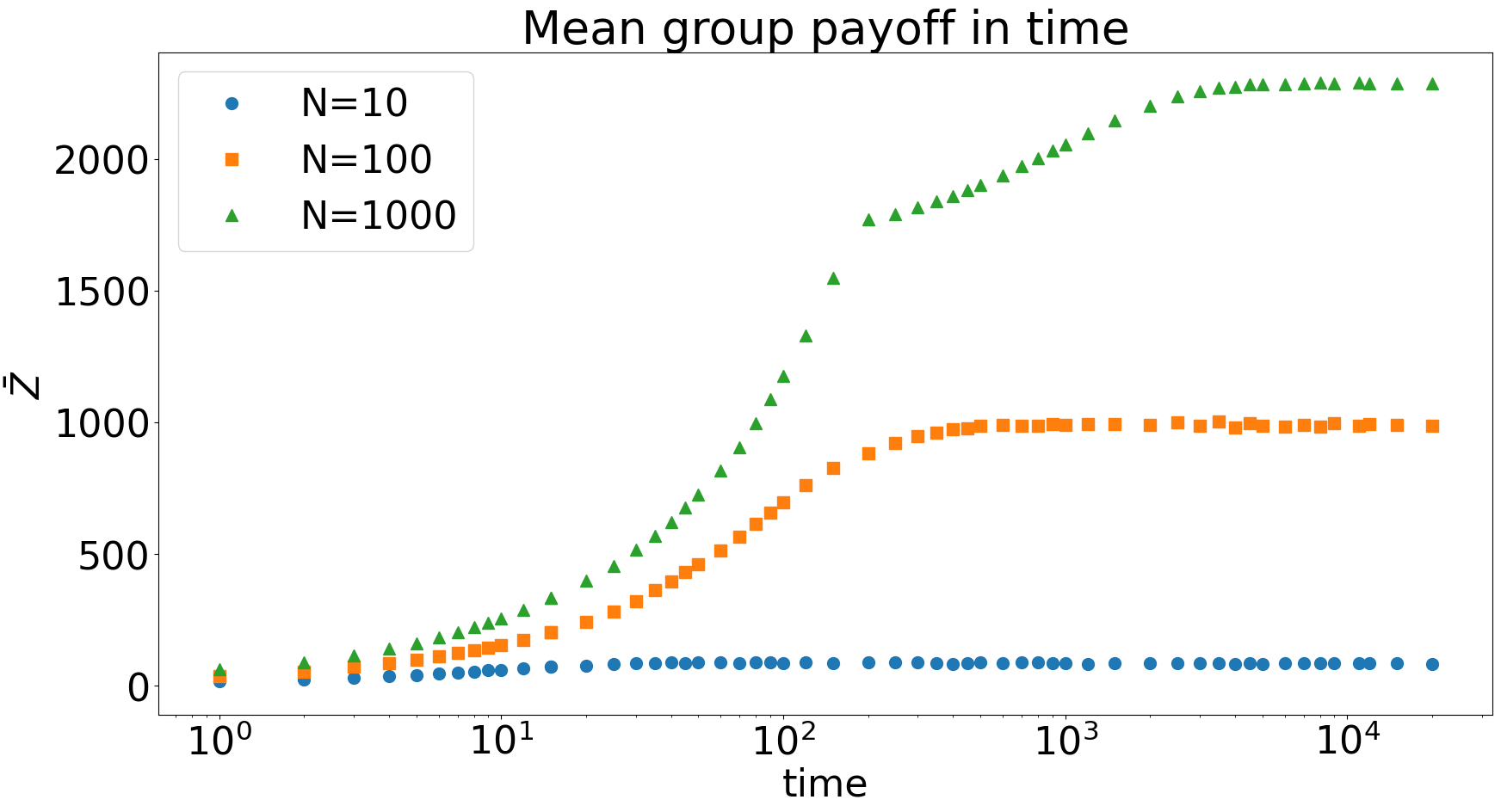}
    \includegraphics[width=0.45\columnwidth]{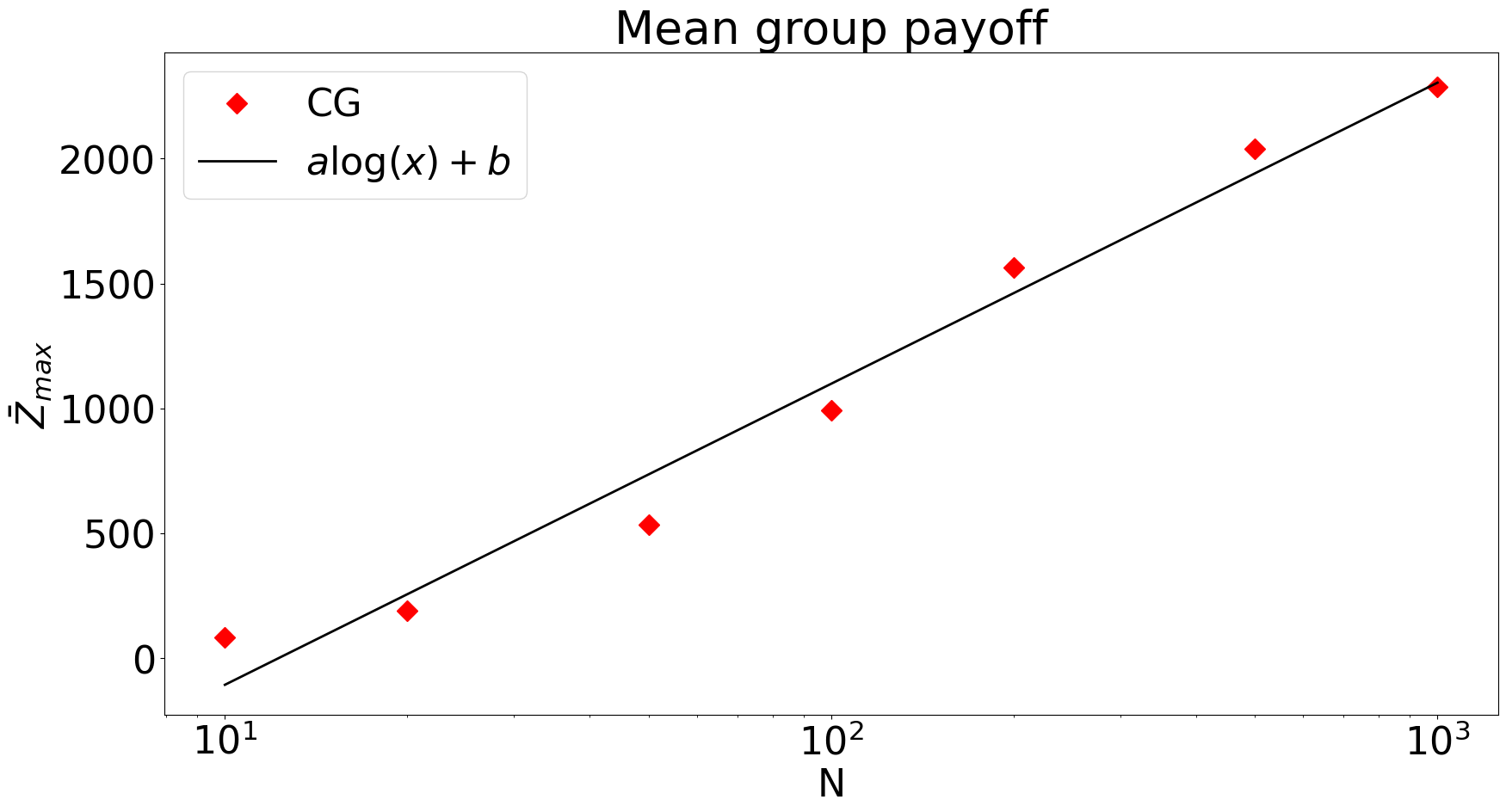}
   
    \caption{(left panel) The mean group payoff for fully connected networks as a function of time for different system sizes. (right panel) Maximally achievable group payoff as a function of system size.  The remaining model parameters are: $B=1000$, $C_i=10$, $C_s=5$, $X=100$.}
    \label{fig:CG}
\end{figure}

Figure \ref{fig:CG} depicts the results for fully connected networks ($p=1$). Similar to the findings reported in \cite{10.1371/journal.pone.0018239}, we observe that after the initial growth of $\bar{Z}$ the dynamics reach a quasi-stationary state where the mean group payoff approaches its maximum value, denoted as $\bar{Z}_\mathrm{max}$. In line with classical results in cultural evolution \cite{Henrich2004}, we find that the maximum group performance increases with the system size uncovering a logarithmic relationship. This relationship becomes more clear by plotting $\bar{Z}_\mathrm{max}$ as a function of  $\log(N)$ (system size), as depicted in the right panel of Figure \ref{fig:CG}.

\begin{figure}[!ht]
    \centering
    \includegraphics[width=2.1in]{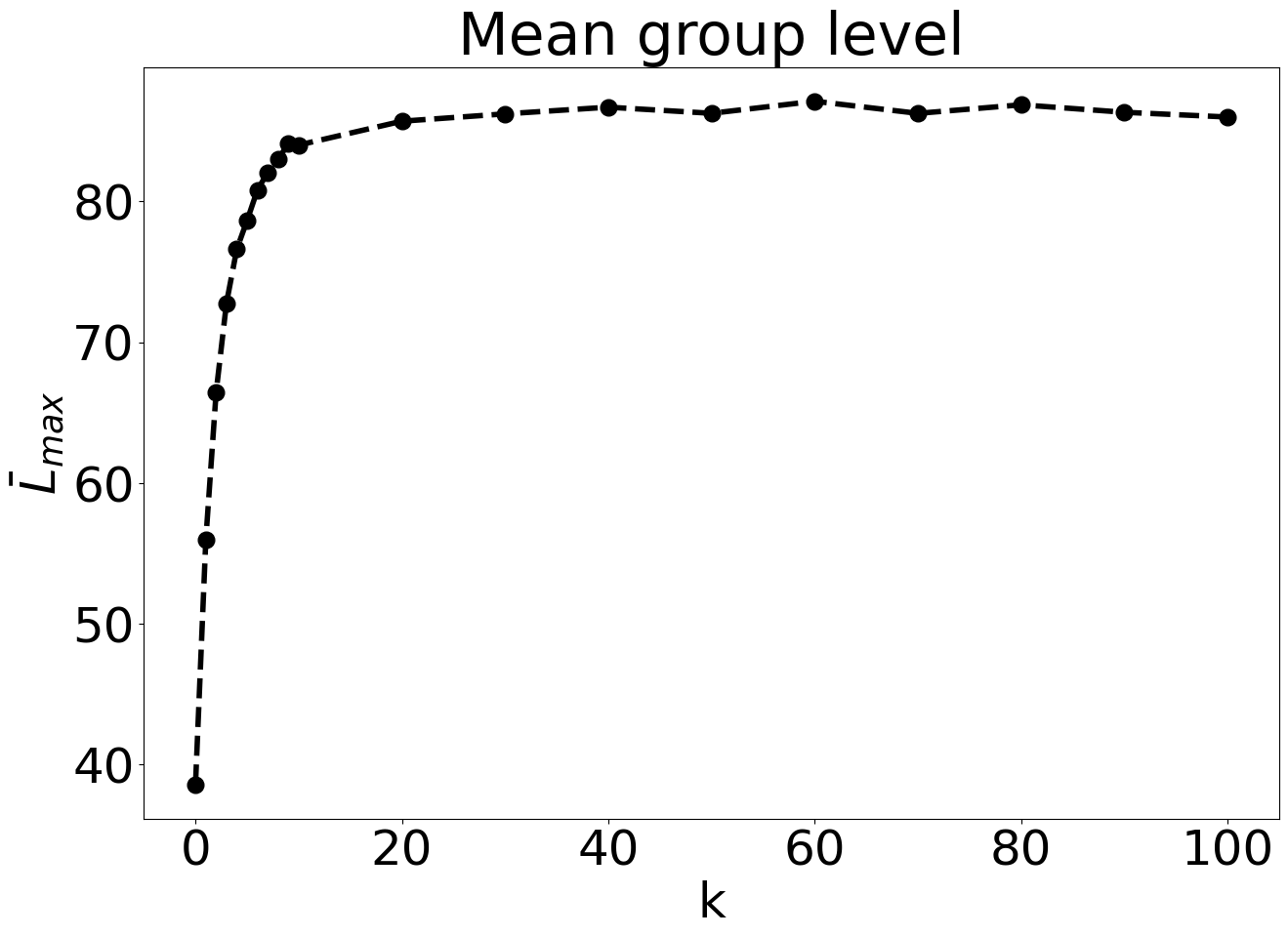}
    \includegraphics[width=2.15in]{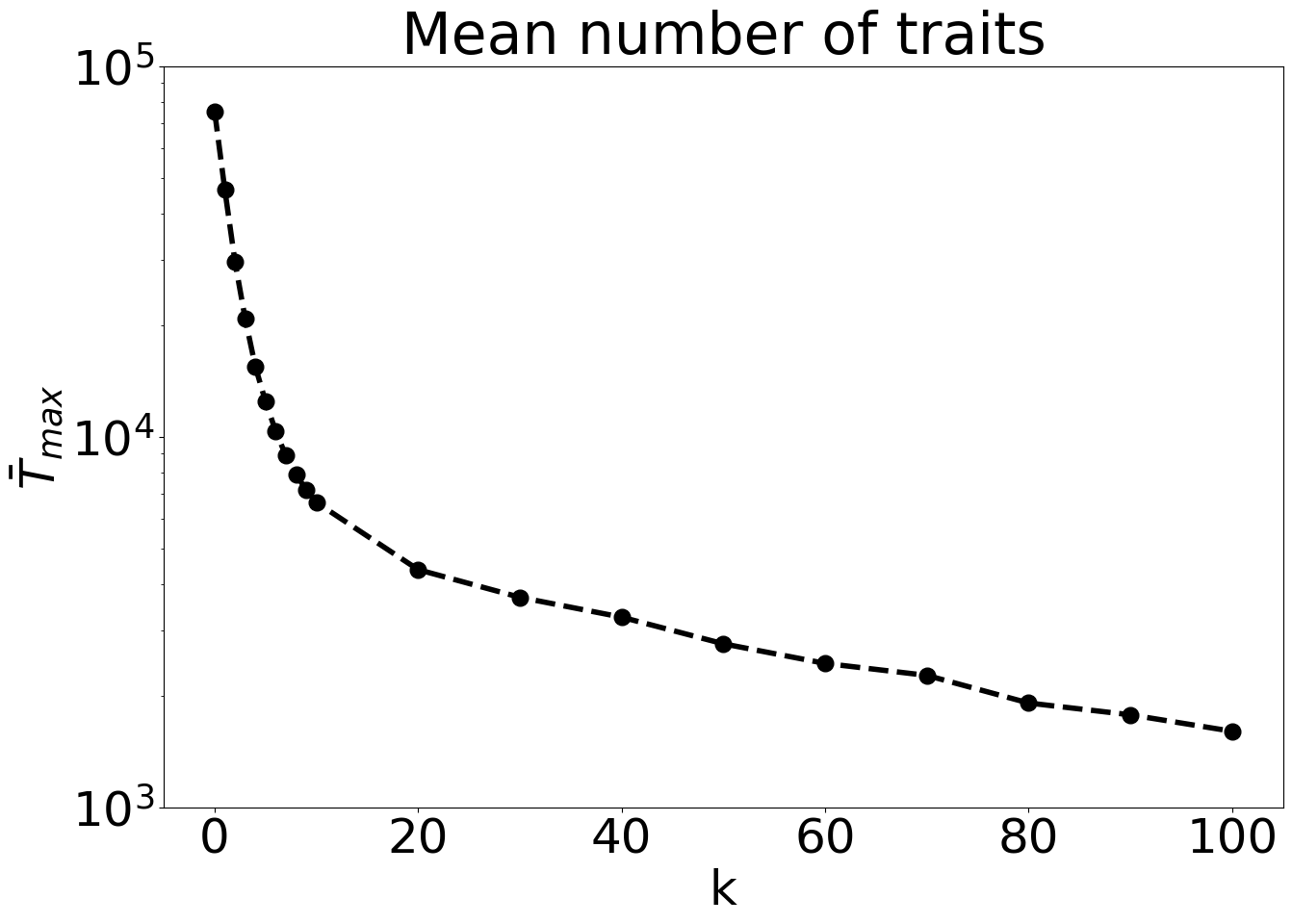} 
    \includegraphics[width=2.2in]{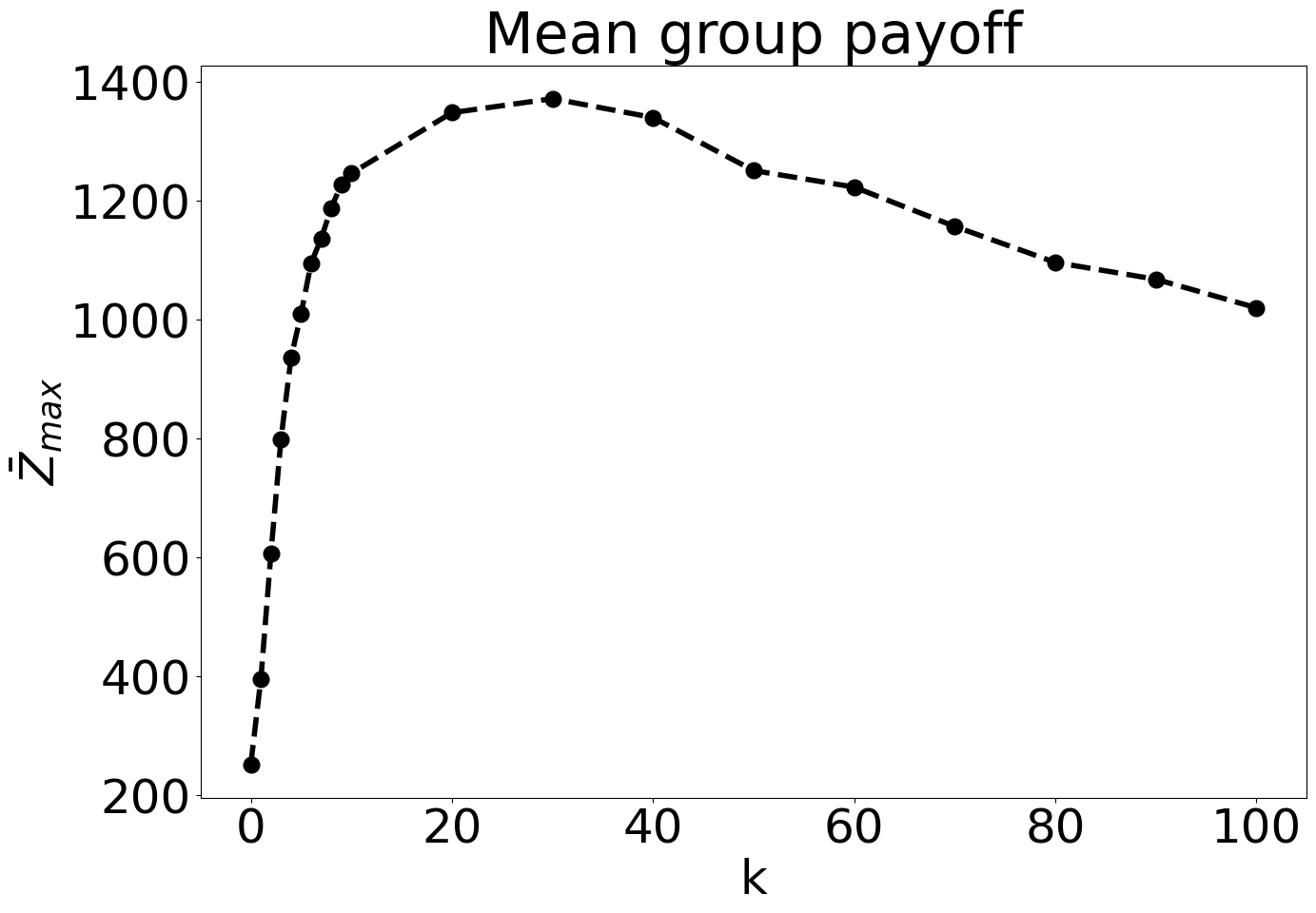}
    \caption{Stationary states as functions of network connectivity $k$. From left to right we plot mean group level $\bar{L}_\mathrm{max}$, mean number of traits $\bar{T}_\mathrm{max}$, and mean group payoff $\bar{Z}_\mathrm{max}$. The remaining model parameters are: $N=100$, $B=1000$, $C_i=10$, $C_s=5$, $X=100$}
    \label{fig:ER}
\end{figure}

Figure \ref{fig:ER} shows the results for random networks with different connectivity, as parameterized by the average number of neighbours $k$ of an agent. 
From left to right we depict the mean group level $\bar{L}_\mathrm{max}$, the mean number of cultural traits $\bar{T}_\mathrm{max}$, and the mean group payoff $\bar{Z}_\mathrm{max}$ as functions of $k$.
The mean group level grows with the number of neighbours until it reaches a plateau for very dense networks. 
Instead, the number of discovered items $\bar{T}_\mathrm{max}$ decreases rapidly and monotonically with $k$. 
The most striking behavior is observed for the mean group payoff. 
Initially, for very sparse networks (low values of $k$) we observe an increase of $\bar{Z}_\mathrm{max}$with $k$, but after a maximum value is reached, for intermediate levels of connectivity, we observe that $\bar{Z}_\mathrm{max}$ decreases with $k$.

\subsection{Algorithmic mediation}
\begin{figure}[!ht]
    \centering
    \includegraphics[width=2.1in]{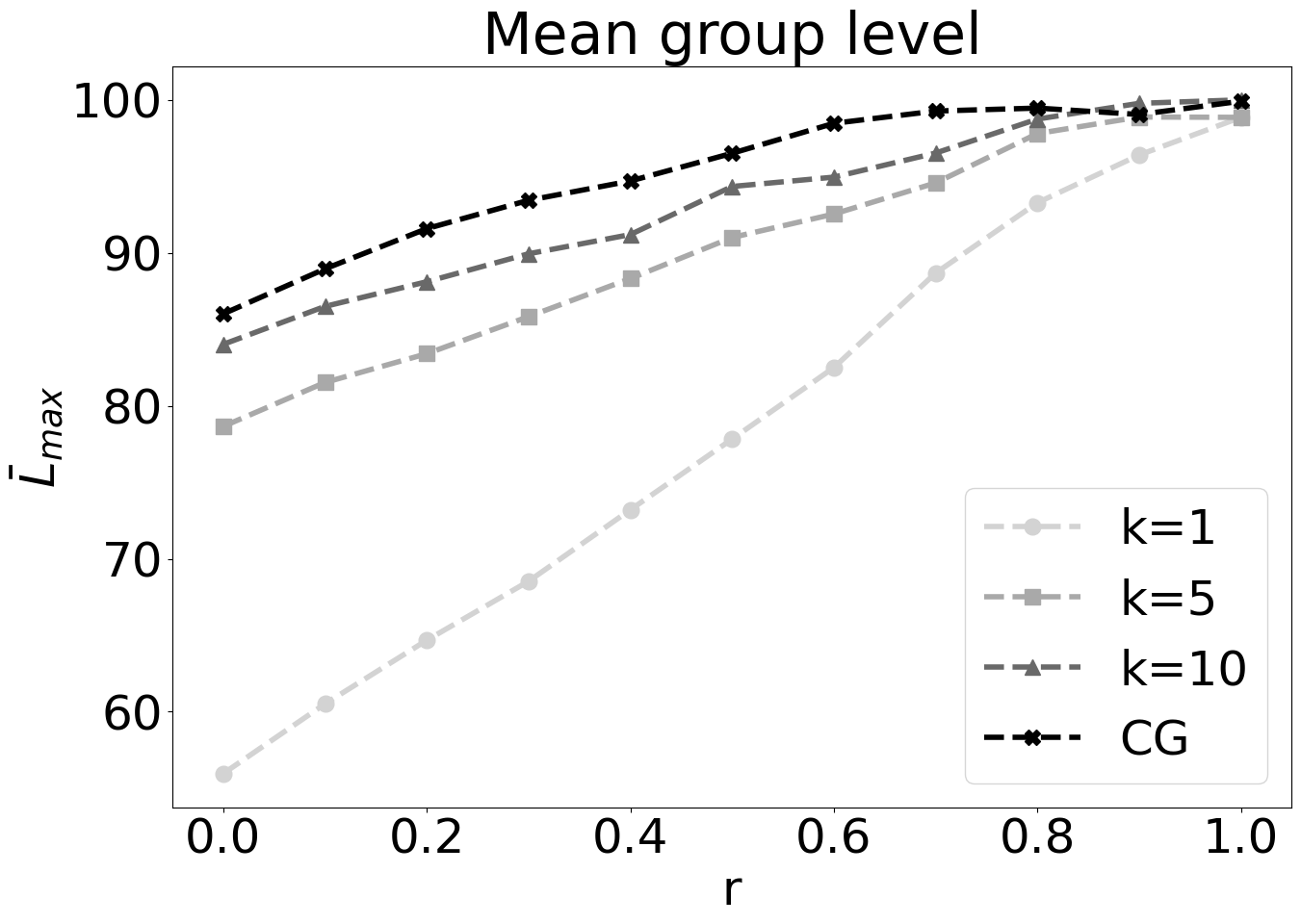}
    \includegraphics[width=2.1in]{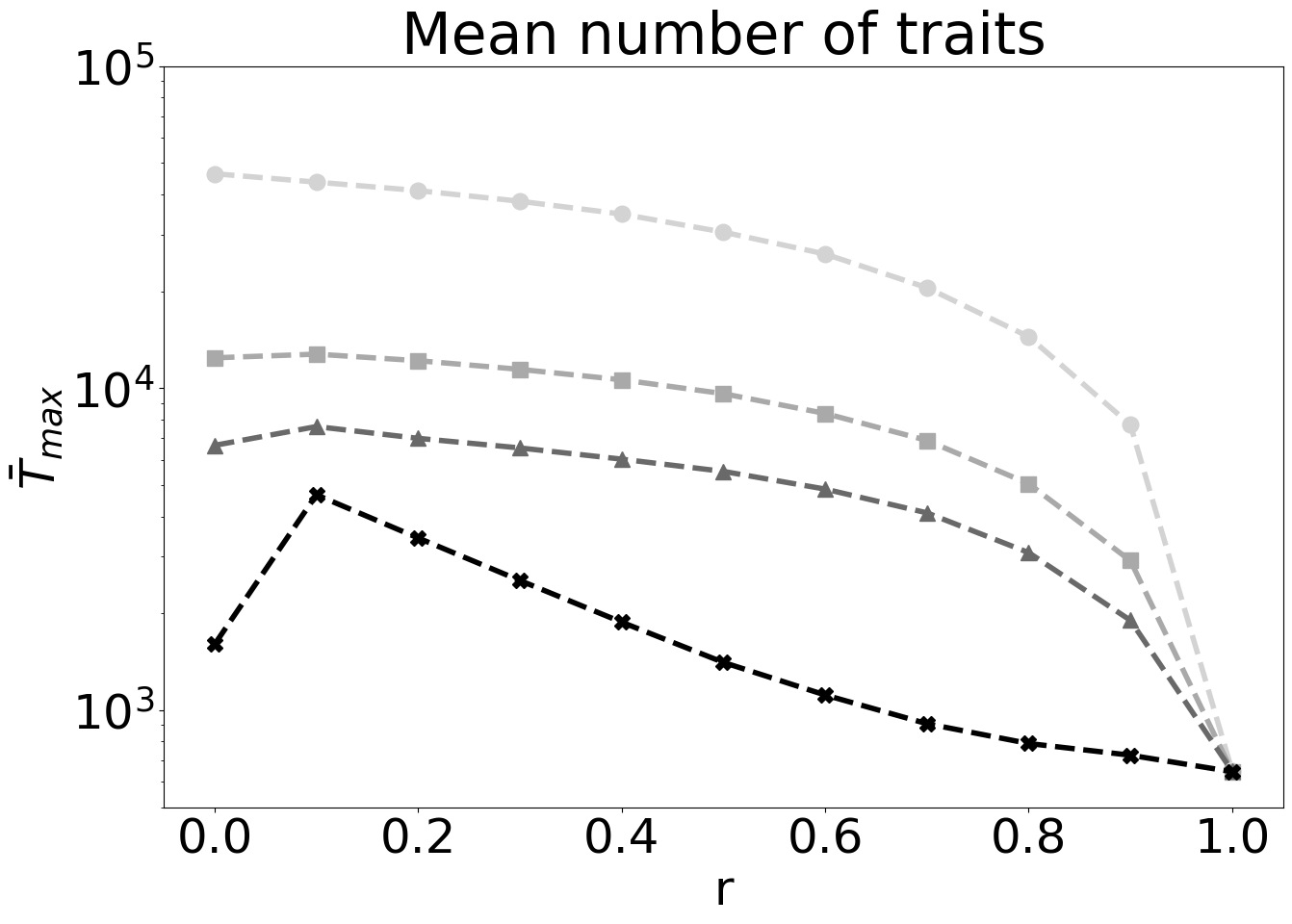} 
    \includegraphics[width=2.1in]{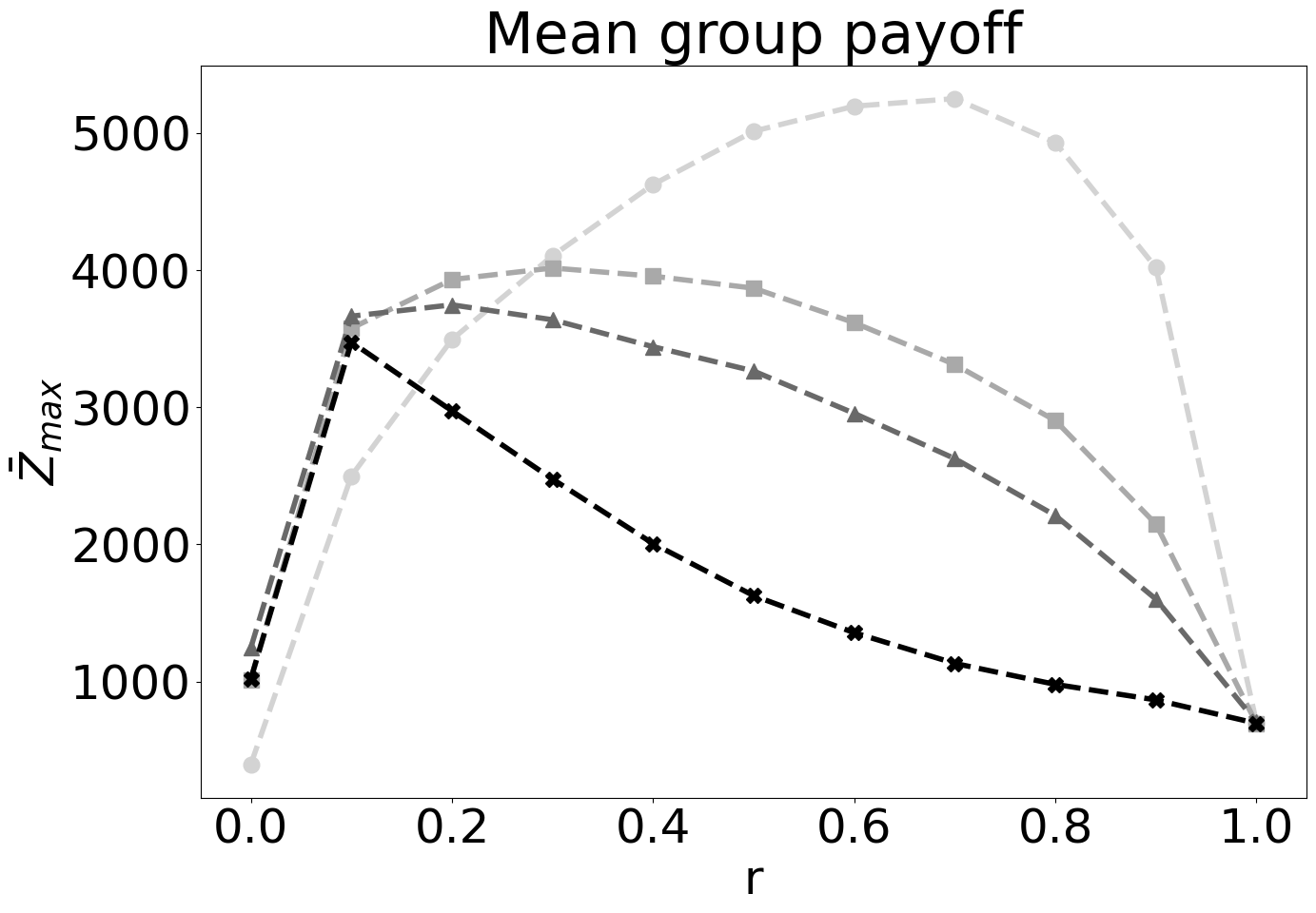}
    \caption{Stationary states for mean group level $\bar{L}_\mathrm{max}$, mean number of traits $\bar{T}_\mathrm{max}$, and mean group payoff $\bar{Z}_\mathrm{max}$ (from left to right) as a function of the ratio of algorithmic mediation  $r$, and for different network connectivities $k$ including complete graphs (CG). The remaining model parameters are: $N=100$, $B=1000$, $C_i=10$, $C_s=5$, $C_r=1$, $X=100$}
    \label{fig:RS}
\end{figure}

\begin{figure}[!ht]
    \centering
    \includegraphics[width=0.75\columnwidth]{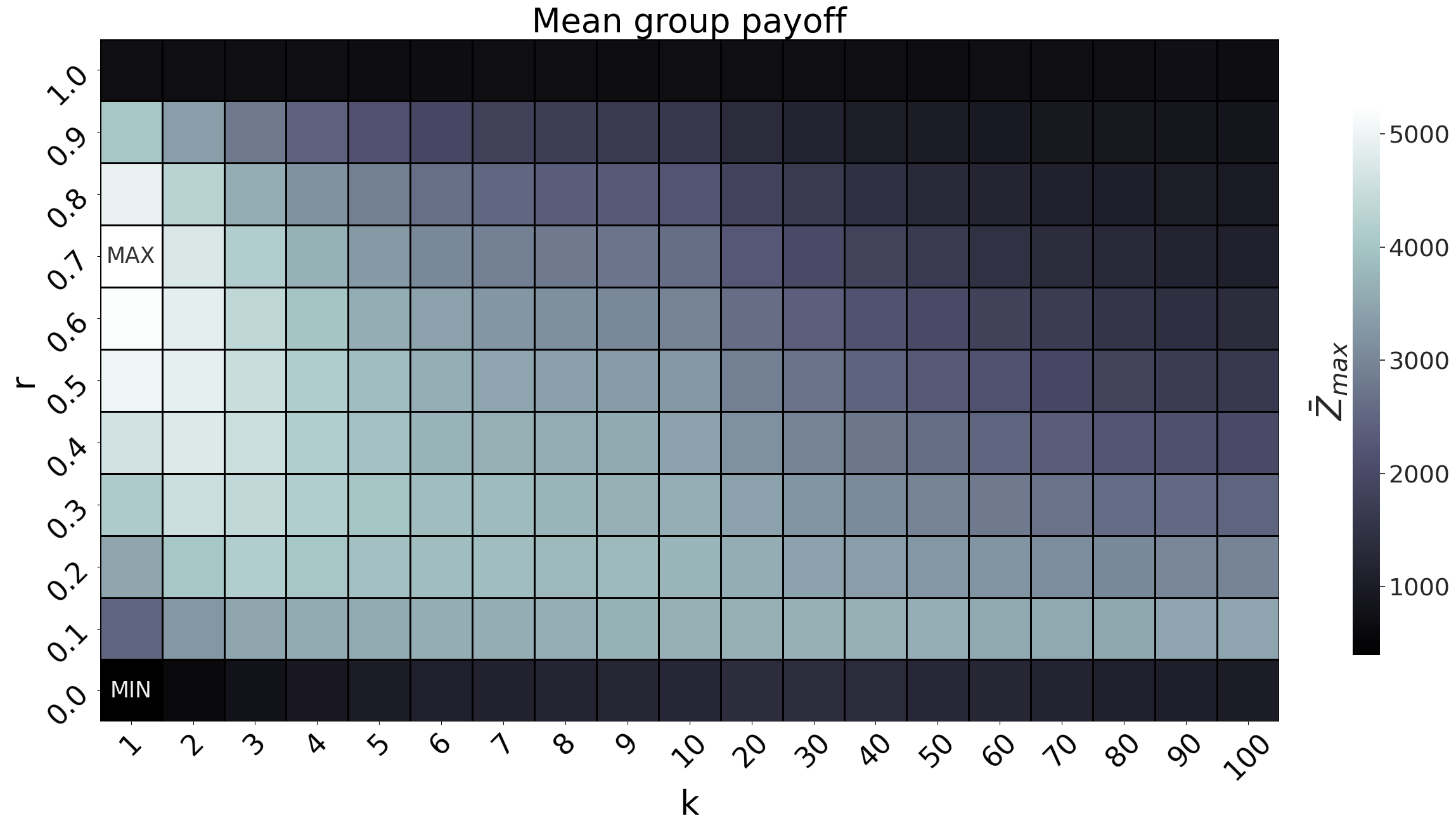} 
    \caption{Heatmap of the mean group payoff $\bar{Z}_\mathrm{max}$ as a function of the average connectivity $k$ and the ratio of algorithmic mediation $r$.  The remaining model parameters are: $N=100$, $B=1000$, $C_i=10$, $C_s=5$, $C_r=1$, $X=100$.}
    \label{fig:RS3d}
\end{figure}

In Fig.~\ref{fig:RS}, we show the results for the impact of algorithmic mediation on cumulative cultural evolution.
Only the mean group level, $\bar{L}_\mathrm{max}$, shows a monotonous increase with increasing ratio of algorithmic mediation $r$ for every value of $k$, see left panel of Fig.~\ref{fig:RS}.   
In line with the observations from the previous section, a higher level of development is observed for denser networks. By contrast, the mean group payoff $\bar{Z}_\mathrm{max}$ gives rise to maximum values for intermediate levels of $r$, and these maximum values decrease with $k$.
This suggests a trade-off between the network connectivity ($k$) and the ratio of algorithmic mediation ($r$), which can be seen more clearly in Fig. \ref{fig:RS3d}. 
The highest payoff is observed for very sparse networks (every agent has only one neighbour on average) and relatively high levels of algorithmic mediation ($r\approx 0.7$), which might be due to the high number of innovated traits $\bar{T}_\mathrm{max}$, for low values of $k$, as can be seen from the bottom panel of Fig.~ \ref{fig:RS}. With fewer neighbours an individual benefits from a higher level of algorithmic mediation, but this trend decreases with $k$.
Note that both limiting cases of $r=0$ and $r=1$ never give rise to maximum values, independently of the connectivity, suggesting that network-based social learning and algorithmic-mediation of information are mutually beneficial. 

From our numerical simulations, it appears that both the size of the system ($N$) and the number of possible development paths ($X$), do not qualitatively affect our modeling results. 
We also did not observe non-trivial relationships between the effort budget ($B$) and the measures of cultural accumulation. Intuitively, a larger effort budget translates into higher payoffs, where the resulting dynamics depend on the network topology and the ratio of algorithmic mediation $r$. In the SM, we report additional simulation results on how the costs of acquiring cultural traits (i.e. $C_i$, $C_s$, $C_r$) affect the dynamics. We find that lower values of the learning costs are associated with higher measures of cultural accumulation, where the impact of ($C_i$, $C_s$, $C_r$) is most pronounced for the mean group level, and much smaller for the mean number of traits and the mean group payoff. We also noticed some qualitative changes of $\bar{Z}_{max}$ as a function of the ratio of algorithmic mediation $r$ for different values of $C_i$, $C_s$, and $C_r$. Importantly, we find that in every investigated case our main result remains unchanged, i.e. intermediate values of $r$ give rise to the maximum performance (peak values of $\bar{Z}_{max}$), where the exact location of that optimum depends on the costs of innovation, social learning and algorithmic mediation, as reported in the SM.

\section{Discussion}
We investigated the effects of network-based social learning and algorithmic mediation on cultural accumulation by building on, and extending, previous modeling efforts in cultural evolution \cite{10.1371/journal.pone.0018239}.  
In the first part, we considered the model without algorithmic mediation and found a strong dependence of cultural accumulation on population size.  
In particular, for fully connected networks, the population's ability to accumulate culture exhibited a positive relationship with population size, characterized by logarithmic growth. This result aligns with established findings in cultural evolution, highlighting the role of population size in fostering cumulative cultural dynamics \cite{henrich2001evolution,lehmann2009coevolution,fogarty2017niche,caldwell2010human,derex2013experimental,muthukrishna2014sociality}.
We then investigated cultural accumulation in random networks that differ in the density of connections, as quantified by the average number of neighbours $k$ of a single agent.  
Although increasing the link density always led to higher levels of development ($\bar{L}_\mathrm{max}$), this was not the case for the mean group payoff ($\bar{Z}_\mathrm{max}$) and the number of innovated traits ($\bar{T}_\mathrm{max}$), which peak at intermediate values of $k$ and monotonously decrease with $k$, respectively.  
These non-trivial behaviors indicate that cultural accumulation is subject to an exploration-exploitation trade-off between innovation (exploration) and social learning (exploitation), a phenomenon that was previously also observed in both experimental and theoretical works on social dynamics including cultural evolution \cite{derex2016partial,smolla2019cultural}, wisdom of the crowds \cite{centola2022network}, as well as collective problem-solving \cite{mason2012collaborative,lazer2007network,barkoczi2016social,baumann2024network}.  

Dense random networks (i.e., networks with high values of $k$) generally allow for the efficient dissemination of information about cultural traits (exploitation) and therefore lead to high cultural complexity. However, this comes at the expense of cultural diversity (number of traits), as the rate of innovation is impaired due to an over-reliance on social learning.  
Furthermore, previous empirical and theoretical findings in cultural evolution and social learning studies suggest that collective performance depends on both the complexity of the task and the underlying network structure. In particular, it was found that performance in dense networks is often decreased for complex tasks \cite{lazer2007network,barkoczi2016social,baumann2024network}, a phenomenon best explained by sufficient levels of transient diversity in less densely connected systems \cite{smaldino2024maintaining}.  
In Fig.~\ref{fig:ER}, the shapes of $\bar{L}_\mathrm{max}$ and $\bar{Z}_\mathrm{max}$, as functions of $k$, therefore indicate that the different sub-processes of cultural accumulation vary in complexity:  
While reaching a high level of complexity is a rather simple problem (high value of $\bar{L}_\mathrm{max}$ in dense networks), it is more complex to collectively coordinate on high average payoffs (low values of $\bar{Z}_\mathrm{max}$ in dense networks).  
In other words, populations develop traits and reach high complexity rather easily, however, such traits do not necessarily have a high utility (i.e. high payoff). In our model, collective payoffs are optimal at a medium level of connectivity, where agents optimally solve the explore-exploit trade-off, i.e., have access to the right amount of information through social learning while maintaining sufficient incentives to innovate.

In the second part, we investigated the combined impact of network-based social learning and algorithmic mediation on cultural accumulation.  
For a fixed value of $k$, we found that increasing $r$—the ratio between algorithmic mediation over network-based social learning—strongly affects the overall dynamics: while the cultural complexity monotonously increases with $r$, the mean payoff gives rise to maximum values for intermediate levels of algorithmic mediation, which suggests a non-trivial interaction between $k$ and $r$. 
The less we are connected, the more we can benefit from algorithmic mediation.  
For a given link density, however, the benefits of increased information sharing via algorithmic mediation become outweighed by their detrimental consequences: too high levels of $r$ create too much effective connectivity, which decreases performance as $r\rightarrow 1$. This result is in line with the strongly decreasing number of traits as $r$ increases, indicating low levels of cultural diversity.  
Interestingly, this behavior remains remarkably stable when the structure of the underlying network is altered, provided the link density is kept constant, as demonstrated in the Supplemental Material (SM). For two additional network structures---small-world networks \cite{watts1998collective} and scale-free networks \cite{barabasi1999emergence}---the mean group level, the number of traits, and the payoffs exhibit the same qualitative behavior as observed in random ER networks.

Furthermore, we examined alternative social learning strategies (in addition to the payoff-biased learning addressed previously) and their influence on cultural accumulation in the presence of algorithmic mediation. Two strategies were considered: one in which cultural traits are copied from the neighbour with the highest complexity level, and alternatively, one in which a random neighbour is chosen. The latter strategy is analogous to the unbiased transmission described in \cite{10.1371/journal.pone.0018239}. It can be seen that for $r = 0$ (solely social learning), the random strategy performs considerably less well than the other two, but the situation becomes much more intricate when algorithmic mediation is introduced (see Figure 2 in SM). The mean number of levels as a function of $r$ is considerably lower for the random selection strategy, yet it yields larger payoffs for high values of $r$. This seemingly contradictory effect is the result of two mechanisms: firstly, a randomly selected neighbour is rarely at a sufficiently high level to allow for copying, which leads to a high number of innovations (as evidenced by the values of the mean number of traits in the middle panel of Figure 2 in  SM); secondly, when the proportion of algorithmic recommendations is sufficiently high, the innovated items are efficiently distributed throughout the system. Consequently, a seemingly not effective random strategy at the individual level becomes optimal for the population.

These results make a clear case for studying the dynamics of algorithmic mediation not in isolation but in relation to social learning strategies motivated by cognitive science (e.g., payoff-biased learning).  
In our model, both algorithmic mediation and social learning acting alone consistently lead to worse payoffs and fewer cultural traits than a combination of both processes.  
This behavior may point toward a general trade-off in socio-technical systems between access to information (and thus the possibility to exchange it) and the ability of individuals to innovate: while algorithms can be beneficial for cultural accumulation, they may also lead to detrimental effects if used excessively \cite{shumailov2024ai,piao2023human}.  

The real-world counterparts of these dynamics provide valuable insights into modern socio-technical systems. For example, recommender systems and large language models (LLMs) shape cultural accumulation by redistributing information globally, facilitating knowledge acquisition and skill development. However, they also risk suppressing diversity or incentivizing conformity when over-relied upon. Applications in collaborative platforms \cite{korkmaz2024github,yasseri2012dynamics}, professional networks \cite{tsagarakis2024analyzing}, or online communities \cite{guess2023social,lucchini2022reddit} could serve as real-world analogs of the processes modeled here. Empirical studies might examine how varying levels of algorithmic support in such environments affect innovation rates or cultural diversity, shedding light on the interplay between cultural exploration and exploitation in digital ecosystems \cite{cui2024ai,brinkmann2023machine,kapoor2021socio,pedreschi2024human}.

Our model has several limitations that should be addressed in future work. First, it focuses exclusively on algorithmic mediation, i.e. the redistribution of existing cultural traits and does not account for algorithmic innovation, where intelligent systems actively generate new solutions or artifacts. Previous research has found strong support for the ability of machines to also take part in the innovation process \cite{shin2023superhuman,sourati2023accelerating}, which future models of cultural evolution should take into account. Second, we assume that algorithmic mediation remains static, whereas the behavior of the real-world algorithms co-evolves with human behavior driven by user feedback and iterative improvement of machines \cite{pedreschi2024human}.
Third, our model does not include biases in algorithmic mediation, such as leading to unintended distortions in content distribution, all of which are prevalent in real systems \cite{kordzadeh2022algorithmic}. Furthermore, extending our framework to account for more complex cultural spaces will be an essential next step for capturing the richness of real-world cultural dynamics, including more complex innovation processes such as recombination \cite{youn2015invention}.
Ultimately, the predictions of our model on the interaction between algorithmic mediation and social learning must stand up to empirical testing through controlled laboratory experiments and observational studies in diverse online and offline environments.

\section*{Authors' Contributions}
A.C.: conceptualization, investigation, software, visualization, writing—original draft, writing—review and editing; 
F.B.: conceptualization, investigation, visualization, writing—original draft, writing—review and editing; 
I.R.: conceptualization, investigation, writing—original draft, writing—review and editing.

All authors gave final approval for publication and agreed to be held accountable for the work performed therein.

\section*{Acknowledgments}
The authors would like to thank the reviewers for their constructive comments.

\section*{Data Accessibility}
Simulation code to reproduce the results in this paper is publicly available in GitHub repository \cite{czaplicka_github} and permanently archived on Zenodo \cite{czaplicka_2025_14718450}.
Electronic supplementary material is available online \cite{czaplicka_baumann_rahwan_2025}.

\section*{Funding Statement}
No additional funding was received for this study.

\bibliography{bibliography}

\begin{thebibliography}{66}
\providecommand{\natexlab}[1]{#1}
\providecommand{\url}[1]{\texttt{#1}}
\expandafter\ifx\csname urlstyle\endcsname\relax
  \providecommand{\doi}[1]{doi: #1}\else
  \providecommand{\doi}{doi: \begingroup \urlstyle{rm}\Url}\fi

\bibitem[Boyd et~al.(1996)Boyd, Richerson, et~al.]{boyd1996culture}
Robert Boyd, Peter~J Richerson, et~al.
\newblock Why culture is common, but cultural evolution is rare.
\newblock In \emph{Proceedings-British Academy}, volume~88, pages 77--94. Oxford University Press Inc., 1996.

\bibitem[Tomasello(2009)]{tomasello2009cultural}
Michael Tomasello.
\newblock \emph{The cultural origins of human cognition}.
\newblock Harvard university press, 2009.

\bibitem[Henrich(2016)]{henrich2016secret}
Joseph Henrich.
\newblock \emph{The secret of our success: How culture is driving human evolution, domesticating our species, and making us smarter}.
\newblock princeton University press, 2016.

\bibitem[Laland(2017)]{laland2017darwin}
Kevin~N Laland.
\newblock \emph{Darwin's unfinished symphony: how culture made the human mind}.
\newblock Princeton University Press, 2017.

\bibitem[Kolodny et~al.(2015)Kolodny, Creanza, and Feldman]{kolodny2015evolution}
Oren Kolodny, Nicole Creanza, and Marcus~W Feldman.
\newblock Evolution in leaps: the punctuated accumulation and loss of cultural innovations.
\newblock \emph{Proceedings of the National Academy of Sciences}, 112\penalty0 (49):\penalty0 E6762--E6769, 2015.

\bibitem[Henrich and Gil-White(2001)]{henrich2001evolution}
Joseph Henrich and Francisco~J Gil-White.
\newblock The evolution of prestige: Freely conferred deference as a mechanism for enhancing the benefits of cultural transmission.
\newblock \emph{Evolution and human behavior}, 22\penalty0 (3):\penalty0 165--196, 2001.

\bibitem[Thompson et~al.(2022)Thompson, Van~Opheusden, Sumers, and Griffiths]{thompson2022complex}
B~Thompson, B~Van~Opheusden, T~Sumers, and TL~Griffiths.
\newblock Complex cognitive algorithms preserved by selective social learning in experimental populations.
\newblock \emph{Science}, 376\penalty0 (6588):\penalty0 95--98, 2022.

\bibitem[Jim{\'e}nez and Mesoudi(2019)]{jimenez2019prestige}
{\'A}ngel~V Jim{\'e}nez and Alex Mesoudi.
\newblock Prestige-biased social learning: Current evidence and outstanding questions.
\newblock \emph{Palgrave Communications}, 5\penalty0 (1), 2019.

\bibitem[Hahn and Bentley(2003)]{hahn2003drift}
Matthew~W Hahn and R~Alexander Bentley.
\newblock Drift as a mechanism for cultural change: an example from baby names.
\newblock \emph{Proceedings of the Royal Society of London. Series B: Biological Sciences}, 270\penalty0 (suppl\_1):\penalty0 S120--S123, 2003.

\bibitem[Herzog et~al.(2004)Herzog, Bentley, and Hahn]{herzog2004random}
Harold~A Herzog, R~Alexander Bentley, and Matthew~W Hahn.
\newblock Random drift and large shifts in popularity of dog breeds.
\newblock \emph{Proceedings of the Royal Society of London. Series B: Biological Sciences}, 271\penalty0 (suppl\_5):\penalty0 S353--S356, 2004.

\bibitem[Mesoudi et~al.(2006)Mesoudi, Whiten, and Dunbar]{mesoudi2006bias}
Alex Mesoudi, Andrew Whiten, and Robin Dunbar.
\newblock A bias for social information in human cultural transmission.
\newblock \emph{British journal of psychology}, 97\penalty0 (3):\penalty0 405--423, 2006.

\bibitem[Mesoudi and Thornton(2018)]{mesoudi2018cumulative}
Alex Mesoudi and Alex Thornton.
\newblock What is cumulative cultural evolution?
\newblock \emph{Proceedings of the Royal Society B}, 285\penalty0 (1880):\penalty0 20180712, 2018.

\bibitem[Henrich(2004)]{Henrich2004}
Joseph Henrich.
\newblock Demography and cultural evolution: How adaptive cultural processes can produce maladaptive losses—the tasmanian case.
\newblock \emph{American Antiquity}, 69\penalty0 (2):\penalty0 197–214, 2004.
\newblock \doi{10.2307/4128416}.

\bibitem[Derex and Boyd(2016)]{derex2016partial}
Maxime Derex and Robert Boyd.
\newblock Partial connectivity increases cultural accumulation within groups.
\newblock \emph{Proceedings of the National Academy of Sciences}, 113\penalty0 (11):\penalty0 2982--2987, 2016.

\bibitem[Miu et~al.(2024)Miu, Rendell, Bowles, Boyd, Cownden, Enquist, Eriksson, Feldman, Lillicrap, McElreath, et~al.]{miu2024refinement}
Elena Miu, Luke Rendell, Sam Bowles, Rob Boyd, Daniel Cownden, Magnus Enquist, Kimmo Eriksson, Marcus~W Feldman, Timothy Lillicrap, Richard McElreath, et~al.
\newblock The refinement paradox and cumulative cultural evolution: Complex products of collective improvement favor conformist outcomes, blind copying, and hyper-credulity.
\newblock \emph{PLOS Computational Biology}, 20\penalty0 (9):\penalty0 e1012436, 2024.

\bibitem[Brinkmann et~al.(2023)Brinkmann, Baumann, Bonnefon, Derex, M{\"u}ller, Nussberger, Czaplicka, Acerbi, Griffiths, Henrich, et~al.]{brinkmann2023machine}
Levin Brinkmann, Fabian Baumann, Jean-Fran{\c{c}}ois Bonnefon, Maxime Derex, Thomas~F M{\"u}ller, Anne-Marie Nussberger, Agnieszka Czaplicka, Alberto Acerbi, Thomas~L Griffiths, Joseph Henrich, et~al.
\newblock Machine culture.
\newblock \emph{Nature Human Behaviour}, 7\penalty0 (11):\penalty0 1855--1868, 2023.

\bibitem[Roy and Dutta(2022)]{roy2022systematic}
Deepjyoti Roy and Mala Dutta.
\newblock A systematic review and research perspective on recommender systems.
\newblock \emph{Journal of Big Data}, 9\penalty0 (1):\penalty0 59, 2022.

\bibitem[Cotter et~al.(2017)Cotter, Cho, and Rader]{cotter2017explaining}
Kelley Cotter, Janghee Cho, and Emilee Rader.
\newblock Explaining the news feed algorithm: An analysis of the" news feed fyi" blog.
\newblock In \emph{Proceedings of the 2017 CHI conference extended abstracts on human factors in computing systems}, pages 1553--1560, 2017.

\bibitem[Page et~al.(1999)Page, Brin, Motwani, and Winograd]{page1999pagerank}
Lawrence Page, Sergey Brin, Rajeev Motwani, and Terry Winograd.
\newblock The pagerank citation ranking: Bringing order to the web.
\newblock Technical report, Stanford infolab, 1999.

\bibitem[Si et~al.(2024)Si, Yang, and Hashimoto]{si2024can}
Chenglei Si, Diyi Yang, and Tatsunori Hashimoto.
\newblock Can llms generate novel research ideas? a large-scale human study with 100+ nlp researchers.
\newblock \emph{arXiv preprint arXiv:2409.04109}, 2024.

\bibitem[Wei et~al.(2022)Wei, Wang, Schuurmans, Bosma, Xia, Chi, Le, Zhou, et~al.]{wei2022chain}
Jason Wei, Xuezhi Wang, Dale Schuurmans, Maarten Bosma, Fei Xia, Ed~Chi, Quoc~V Le, Denny Zhou, et~al.
\newblock Chain-of-thought prompting elicits reasoning in large language models.
\newblock \emph{Advances in neural information processing systems}, 35:\penalty0 24824--24837, 2022.

\bibitem[Kazemitabaar et~al.(2023)Kazemitabaar, Hou, Henley, Ericson, Weintrop, and Grossman]{kazemitabaar2023novices}
Majeed Kazemitabaar, Xinying Hou, Austin Henley, Barbara~Jane Ericson, David Weintrop, and Tovi Grossman.
\newblock How novices use llm-based code generators to solve cs1 coding tasks in a self-paced learning environment.
\newblock In \emph{Proceedings of the 23rd Koli Calling International Conference on Computing Education Research}, pages 1--12, 2023.

\bibitem[Chevalier et~al.(2024)Chevalier, Geng, Wettig, Chen, Mizera, Annala, Aragon, Fanlo, Frieder, Machado, et~al.]{chevalier2024language}
Alexis Chevalier, Jiayi Geng, Alexander Wettig, Howard Chen, Sebastian Mizera, Toni Annala, Max~Jameson Aragon, Arturo~Rodr{\'\i}guez Fanlo, Simon Frieder, Simon Machado, et~al.
\newblock Language models as science tutors.
\newblock \emph{arXiv preprint arXiv:2402.11111}, 2024.

\bibitem[Cockburn et~al.(2018)Cockburn, Henderson, and Stern]{cockburn2018impact}
Iain~M Cockburn, Rebecca Henderson, and Scott Stern.
\newblock \emph{The impact of artificial intelligence on innovation}, volume 24449.
\newblock National bureau of economic research Cambridge, MA, USA, 2018.

\bibitem[Pariser(2011)]{pariser2011filter}
Eli Pariser.
\newblock \emph{The filter bubble: How the new personalized web is changing what we read and how we think}.
\newblock Penguin, 2011.

\bibitem[McPherson et~al.(2001)McPherson, Smith-Lovin, and Cook]{mcpherson2001birds}
Miller McPherson, Lynn Smith-Lovin, and James~M Cook.
\newblock Birds of a feather: Homophily in social networks.
\newblock \emph{Annual review of sociology}, 27\penalty0 (1):\penalty0 415--444, 2001.

\bibitem[Baumann et~al.(2024{\natexlab{a}})Baumann, Halpern, Procaccia, Rahwan, Shapira, and W\"{u}thrich]{10.1145/3589334.3645713}
Fabian Baumann, Daniel Halpern, Ariel~D. Procaccia, Iyad Rahwan, Itai Shapira, and Manuel W\"{u}thrich.
\newblock Optimal engagement-diversity tradeoffs in social media.
\newblock In \emph{Proceedings of the ACM Web Conference 2024}, WWW '24, page 288–299, New York, NY, USA, 2024{\natexlab{a}}. Association for Computing Machinery.
\newblock ISBN 9798400701719.
\newblock \doi{10.1145/3589334.3645713}.
\newblock URL \url{https://doi.org/10.1145/3589334.3645713}.

\bibitem[Perra and Rocha(2019)]{perra2019modelling}
Nicola Perra and Luis~EC Rocha.
\newblock Modelling opinion dynamics in the age of algorithmic personalisation.
\newblock \emph{Scientific reports}, 9\penalty0 (1):\penalty0 7261, 2019.

\bibitem[Lanzetti et~al.(2023)Lanzetti, Dörfler, and Pagan]{lanzetti2023impactrecommendationsystemsopinion}
Nicolas Lanzetti, Florian Dörfler, and Nicolò Pagan.
\newblock The impact of recommendation systems on opinion dynamics: Microscopic versus macroscopic effects, 2023.
\newblock URL \url{https://arxiv.org/abs/2309.08967}.

\bibitem[Guess et~al.(2023)Guess, Malhotra, Pan, Barber{\'a}, Allcott, Brown, Crespo-Tenorio, Dimmery, Freelon, Gentzkow, et~al.]{guess2023social}
Andrew~M Guess, Neil Malhotra, Jennifer Pan, Pablo Barber{\'a}, Hunt Allcott, Taylor Brown, Adriana Crespo-Tenorio, Drew Dimmery, Deen Freelon, Matthew Gentzkow, et~al.
\newblock How do social media feed algorithms affect attitudes and behavior in an election campaign?
\newblock \emph{Science}, 381\penalty0 (6656):\penalty0 398--404, 2023.

\bibitem[Mesoudi(2011)]{10.1371/journal.pone.0018239}
Alex Mesoudi.
\newblock Variable cultural acquisition costs constrain cumulative cultural evolution.
\newblock \emph{PLOS ONE}, 6\penalty0 (3):\penalty0 1--10, 03 2011.
\newblock \doi{10.1371/journal.pone.0018239}.
\newblock URL \url{https://doi.org/10.1371/journal.pone.0018239}.

\bibitem[Rendell et~al.(2010)Rendell, Boyd, Cownden, Enquist, Eriksson, Feldman, Fogarty, Ghirlanda, Lillicrap, and Laland]{Rendell208}
L.~Rendell, R.~Boyd, D.~Cownden, M.~Enquist, K.~Eriksson, M.~W. Feldman, L.~Fogarty, S.~Ghirlanda, T.~Lillicrap, and K.~N. Laland.
\newblock Why copy others? insights from the social learning strategies tournament.
\newblock \emph{Science}, 328\penalty0 (5975):\penalty0 208--213, 2010.
\newblock ISSN 0036-8075.
\newblock \doi{10.1126/science.1184719}.
\newblock URL \url{https://science.sciencemag.org/content/328/5975/208}.

\bibitem[Smolla and Ak{\c{c}}ay(2019)]{smolla2019cultural}
Marco Smolla and Erol Ak{\c{c}}ay.
\newblock Cultural selection shapes network structure.
\newblock \emph{Science advances}, 5\penalty0 (8):\penalty0 eaaw0609, 2019.

\bibitem[Derex et~al.(2018)Derex, Perreault, and Boyd]{derex2018divide}
Maxime Derex, Charles Perreault, and Robert Boyd.
\newblock Divide and conquer: intermediate levels of population fragmentation maximize cultural accumulation.
\newblock \emph{Philosophical Transactions of the Royal Society B: Biological Sciences}, 373\penalty0 (1743):\penalty0 20170062, 2018.

\bibitem[Boyd and Richerson(1988)]{boyd1988culture}
R.~Boyd and P.J. Richerson.
\newblock \emph{Culture and the Evolutionary Process}.
\newblock Biology, Anthropology, Sociology. University of Chicago Press, 1988.
\newblock ISBN 9780226069333.
\newblock URL \url{https://books.google.pl/books?id=MBg4oBsCKU8C}.

\bibitem[Lazer and Friedman(2007)]{lazer2007network}
David Lazer and Allan Friedman.
\newblock The network structure of exploration and exploitation.
\newblock \emph{Administrative science quarterly}, 52\penalty0 (4):\penalty0 667--694, 2007.

\bibitem[Henrich and McElreath(2003)]{henrich2003evolution}
Joseph Henrich and Richard McElreath.
\newblock The evolution of cultural evolution.
\newblock \emph{Evolutionary Anthropology: Issues, News, and Reviews: Issues, News, and Reviews}, 12\penalty0 (3):\penalty0 123--135, 2003.

\bibitem[Park et~al.(2023)Park, Leahey, and Funk]{park2023papers}
Michael Park, Erin Leahey, and Russell~J Funk.
\newblock Papers and patents are becoming less disruptive over time.
\newblock \emph{Nature}, 613\penalty0 (7942):\penalty0 138--144, 2023.

\bibitem[Youn et~al.(2015)Youn, Strumsky, Bettencourt, and Lobo]{youn2015invention}
Hyejin Youn, Deborah Strumsky, Luis~MA Bettencourt, and Jos{\'e} Lobo.
\newblock Invention as a combinatorial process: evidence from us patents.
\newblock \emph{Journal of the Royal Society interface}, 12\penalty0 (106):\penalty0 20150272, 2015.

\bibitem[Lehmann and Feldman(2009)]{lehmann2009coevolution}
Laurent Lehmann and Marcus~W Feldman.
\newblock Coevolution of adaptive technology, maladaptive culture and population size in a producer--scrounger game.
\newblock \emph{Proceedings of the Royal Society B: Biological Sciences}, 276\penalty0 (1674):\penalty0 3853--3862, 2009.

\bibitem[Fogarty and Creanza(2017)]{fogarty2017niche}
Laurel Fogarty and Nicole Creanza.
\newblock The niche construction of cultural complexity: interactions between innovations, population size and the environment.
\newblock \emph{Philosophical Transactions of the Royal Society B: Biological Sciences}, 372\penalty0 (1735):\penalty0 20160428, 2017.

\bibitem[Caldwell and Millen(2010)]{caldwell2010human}
Christine~A Caldwell and Ailsa~E Millen.
\newblock Human cumulative culture in the laboratory: effects of (micro) population size.
\newblock \emph{Learning \& Behavior}, 38\penalty0 (3):\penalty0 310--318, 2010.

\bibitem[Derex et~al.(2013)Derex, Beugin, Godelle, and Raymond]{derex2013experimental}
Maxime Derex, Marie-Pauline Beugin, Bernard Godelle, and Michel Raymond.
\newblock Experimental evidence for the influence of group size on cultural complexity.
\newblock \emph{Nature}, 503\penalty0 (7476):\penalty0 389--391, 2013.

\bibitem[Muthukrishna et~al.(2014)Muthukrishna, Shulman, Vasilescu, and Henrich]{muthukrishna2014sociality}
Michael Muthukrishna, Ben~W Shulman, Vlad Vasilescu, and Joseph Henrich.
\newblock Sociality influences cultural complexity.
\newblock \emph{Proceedings of the Royal Society B: Biological Sciences}, 281\penalty0 (1774):\penalty0 20132511, 2014.

\bibitem[Centola(2022)]{centola2022network}
Damon Centola.
\newblock The network science of collective intelligence.
\newblock \emph{Trends in Cognitive Sciences}, 26\penalty0 (11):\penalty0 923--941, 2022.

\bibitem[Mason and Watts(2012)]{mason2012collaborative}
Winter Mason and Duncan~J Watts.
\newblock Collaborative learning in networks.
\newblock \emph{Proceedings of the National Academy of Sciences}, 109\penalty0 (3):\penalty0 764--769, 2012.

\bibitem[Barkoczi and Galesic(2016)]{barkoczi2016social}
Daniel Barkoczi and Mirta Galesic.
\newblock Social learning strategies modify the effect of network structure on group performance.
\newblock \emph{Nature communications}, 7\penalty0 (1):\penalty0 1--8, 2016.

\bibitem[Baumann et~al.(2024{\natexlab{b}})Baumann, Czaplicka, and Rahwan]{baumann2024network}
Fabian Baumann, Agnieszka Czaplicka, and Iyad Rahwan.
\newblock Network structure shapes the impact of diversity in collective learning.
\newblock \emph{Scientific Reports}, 14\penalty0 (1):\penalty0 2491, 2024{\natexlab{b}}.

\bibitem[Smaldino et~al.(2024)Smaldino, Moser, Perez~Velilla, and Werling]{smaldino2024maintaining}
Paul~E Smaldino, Cody Moser, Alejandro Perez~Velilla, and Mikkel Werling.
\newblock Maintaining transient diversity is a general principle for improving collective problem solving.
\newblock \emph{Perspectives on Psychological Science}, 19\penalty0 (2):\penalty0 454--464, 2024.

\bibitem[Watts and Strogatz(1998)]{watts1998collective}
Duncan~J Watts and Steven~H Strogatz.
\newblock Collective dynamics of ‘small-world’networks.
\newblock \emph{nature}, 393\penalty0 (6684):\penalty0 440--442, 1998.

\bibitem[Barab{\'a}si and Albert(1999)]{barabasi1999emergence}
Albert-L{\'a}szl{\'o} Barab{\'a}si and R{\'e}ka Albert.
\newblock Emergence of scaling in random networks.
\newblock \emph{science}, 286\penalty0 (5439):\penalty0 509--512, 1999.

\bibitem[Shumailov et~al.(2024)Shumailov, Shumaylov, Zhao, Papernot, Anderson, and Gal]{shumailov2024ai}
Ilia Shumailov, Zakhar Shumaylov, Yiren Zhao, Nicolas Papernot, Ross Anderson, and Yarin Gal.
\newblock Ai models collapse when trained on recursively generated data.
\newblock \emph{Nature}, 631\penalty0 (8022):\penalty0 755--759, 2024.

\bibitem[Piao et~al.(2023)Piao, Liu, Zhang, Su, and Li]{piao2023human}
Jinghua Piao, Jiazhen Liu, Fang Zhang, Jun Su, and Yong Li.
\newblock Human--ai adaptive dynamics drives the emergence of information cocoons.
\newblock \emph{Nature Machine Intelligence}, 5\penalty0 (11):\penalty0 1214--1224, 2023.

\bibitem[Korkmaz et~al.(2024)Korkmaz, Calder{\'o}n, Kramer, Guci, and Robbins]{korkmaz2024github}
Gizem Korkmaz, J~Bayo{\'a}n~Santiago Calder{\'o}n, Brandon~L Kramer, Ledia Guci, and Carol~A Robbins.
\newblock From github to gdp: A framework for measuring open source software innovation.
\newblock \emph{Research Policy}, 53\penalty0 (3):\penalty0 104954, 2024.

\bibitem[Yasseri et~al.(2012)Yasseri, Sumi, Rung, Kornai, and Kert{\'e}sz]{yasseri2012dynamics}
Taha Yasseri, Robert Sumi, Andr{\'a}s Rung, Andr{\'a}s Kornai, and J{\'a}nos Kert{\'e}sz.
\newblock Dynamics of conflicts in wikipedia.
\newblock \emph{PloS one}, 7\penalty0 (6):\penalty0 e38869, 2012.

\bibitem[Tsagarakis et~al.(2024)Tsagarakis, Daglis, Gkillas, and Mavragani]{tsagarakis2024analyzing}
Konstantinos~P Tsagarakis, Theodoros Daglis, Konstantinos Gkillas, and Amaryllis Mavragani.
\newblock Analyzing linkedin data to explore the relationships between sustainable development goals, circular economy, and electoral dynamics.
\newblock \emph{Scientific Reports}, 14\penalty0 (1):\penalty0 1--14, 2024.

\bibitem[Lucchini et~al.(2022)Lucchini, Aiello, Alessandretti, De~Francisci~Morales, Starnini, and Baronchelli]{lucchini2022reddit}
Lorenzo Lucchini, Luca~Maria Aiello, Laura Alessandretti, Gianmarco De~Francisci~Morales, Michele Starnini, and Andrea Baronchelli.
\newblock From reddit to wall street: The role of committed minorities in financial collective action.
\newblock \emph{Royal Society Open Science}, 9\penalty0 (4):\penalty0 211488, 2022.

\bibitem[Cui and Yasseri(2024)]{cui2024ai}
Hao Cui and Taha Yasseri.
\newblock Ai-enhanced collective intelligence.
\newblock \emph{Patterns}, 5\penalty0 (11), 2024.

\bibitem[Kapoor et~al.(2021)Kapoor, Bigdeli, Dwivedi, Schroeder, Beltagui, and Baines]{kapoor2021socio}
Kawaljeet Kapoor, Ali~Ziaee Bigdeli, Yogesh~K Dwivedi, Andreas Schroeder, Ahmad Beltagui, and Tim Baines.
\newblock A socio-technical view of platform ecosystems: Systematic review and research agenda.
\newblock \emph{Journal of Business Research}, 128:\penalty0 94--108, 2021.

\bibitem[Pedreschi et~al.(2024)Pedreschi, Pappalardo, Ferragina, Baeza-Yates, Barab{\'a}si, Dignum, Dignum, Eliassi-Rad, Giannotti, Kert{\'e}sz, et~al.]{pedreschi2024human}
Dino Pedreschi, Luca Pappalardo, Emanuele Ferragina, Ricardo Baeza-Yates, Albert-L{\'a}szl{\'o} Barab{\'a}si, Frank Dignum, Virginia Dignum, Tina Eliassi-Rad, Fosca Giannotti, J{\'a}nos Kert{\'e}sz, et~al.
\newblock Human-ai coevolution.
\newblock \emph{Artificial Intelligence}, page 104244, 2024.

\bibitem[Shin et~al.(2023)Shin, Kim, Van~Opheusden, and Griffiths]{shin2023superhuman}
Minkyu Shin, Jin Kim, Bas Van~Opheusden, and Thomas~L Griffiths.
\newblock Superhuman artificial intelligence can improve human decision-making by increasing novelty.
\newblock \emph{Proceedings of the National Academy of Sciences}, 120\penalty0 (12):\penalty0 e2214840120, 2023.

\bibitem[Sourati and Evans(2023)]{sourati2023accelerating}
Jamshid Sourati and James~A Evans.
\newblock Accelerating science with human-aware artificial intelligence.
\newblock \emph{Nature human behaviour}, 7\penalty0 (10):\penalty0 1682--1696, 2023.

\bibitem[Kordzadeh and Ghasemaghaei(2022)]{kordzadeh2022algorithmic}
Nima Kordzadeh and Maryam Ghasemaghaei.
\newblock Algorithmic bias: review, synthesis, and future research directions.
\newblock \emph{European Journal of Information Systems}, 31\penalty0 (3):\penalty0 388--409, 2022.

\bibitem[Czaplicka(2025{\natexlab{a}})]{czaplicka_github}
Agnieszka Czaplicka.
\newblock Mutual benefits of social learning and algorithmic mediation for cumulative culture.
\newblock \url{https://github.com/aga-cz/Mutual-benefits-of-SL-and-RS}, 2025{\natexlab{a}}.

\bibitem[Czaplicka(2025{\natexlab{b}})]{czaplicka_2025_14718450}
Agnieszka Czaplicka.
\newblock Mutual benefits of social learning and algorithmic mediation for cumulative culture, January 2025{\natexlab{b}}.
\newblock URL \url{https://doi.org/10.5281/zenodo.14718450}.

\bibitem[Czaplicka et~al.(2025)Czaplicka, Baumann, and Rahwan]{czaplicka_baumann_rahwan_2025}
Agnieszka Czaplicka, Fabian Baumann, and Iyad Rahwan.
\newblock Supplementary material from \textit{Mutual benefits of social learning and algorithmic mediation for cumulative culture}.
\newblock \url{https://doi.org/10.6084/m9.figshare.c.7718239.v1}, March 2025.

\end{thebibliography}
\bibliographystyle{unsrtnat}

\end{document}